\documentclass[12pt]{article}
\usepackage{amssymb}
\usepackage{amsfonts}
\usepackage{amsmath,amsthm}
\usepackage{graphicx,epsfig }
\usepackage{graphics,epsfig}
\usepackage{subfigure}
\usepackage[symbol]{footmisc}
\usepackage{tikz}
\usetikzlibrary{decorations.pathmorphing}
\usetikzlibrary{shadows,shapes,arrows,chains,decorations.pathreplacing,calc}

\oddsidemargin 10mm \numberwithin{equation}{section}
\def\be{\begin{equation}}
\def\ee{\end{equation}}
\begin{document}
\begin{center} {{\bf {Complexity growth and shock wave geometry in AdS-Maxwell-power-Yang-Mills theory}}
 \vskip 0.50 cm
  {{ Emad Yaraie \footnote{E-mail: eyaraie@semnan.ac.ir
 } }{ Hossein Ghaffarnejad \footnote{E-mail: hghafarnejad@semnan.ac.ir
 } }{ Mohammad Farsam \footnote{E-mail: mhdfarsam@semnan.ac.ir
 } }}\vskip 0.2 cm \textit{Faculty of Physics, Semnan
University, P.C. 35131-19111, Semnan, Iran }}
\end{center}
\begin{abstract}
We study effects of non-abelian gauge fields on the holographic
characteristics for instance the evolution of computational
complexity. To do so we choose Maxwell-power-Yang-Mills theory
defined in the AdS space-time. Then we seek the impact of charge
of the YM field on the complexity growth rate by using
$complexity=action$ (CA) conjecture. We
 also investigate the spreading of perturbations near the horizon and the complexity growth rate in local shock wave geometry in presence of the
 YM charge. At last we check validity regime of Lloyd bound.
\end{abstract}

\section{Introduction}

In the context of AdS/CFT duality a thermal system on field theory could be expressed by a gravity model
 in AdS spacetime. By considering various models of black hole solution in the bulk we can explore field theory behaviors
 which may be complicated when they are studied in a quantum field theory. One of the important aspects of field theory is the computational
  complexity. This is number of qubit gates in the smallest quantum circuit [1,2] or in the another definition is the minimal depth of a
  quantum circuit [3]. Actually computational complexity regarding AdS/CFT duality could explain something about the inside of
  black hole and corresponding geometry. There are two conjectures that relate complexity on the boundary to the geometry of bulk:
  (I) At the first and older one, complexity is supposed to be equal to the maximal volume in the spacelike slice into the bulk [4],
   or CV as
\begin{equation}
\mathcal{C}(t_L,t_R)\sim\frac{V}{L},
\end{equation}
where $G$ is the Newton`s coupling constant, $L$ is a length scale
of the system which is given by the AdS radius for large black
holes and the horizon radius for small black holes. $V$ is the
volume of spacelike slice or the Einstein-Rosen bridge (ER) with
connected points $t_L$ and $t_R$ corresponding to the left and the
right boundaries, respectively. (II) At the newer conjecture,
quantum complexity is proportional to classical action in the bulk
which is defined in "Wheeler-DeWitt" (WD) patch, or CA [5,6]. The
privilege of this conjecture rather than the older one is
needlessness to any length scale chosen by hand, such as "L" or
the event horizon radius,
\begin{equation}
\mathcal{C}(\Sigma)=\frac{\mathcal{A}_{WDW}}{\pi\hbar},
\end{equation}
in which $\Sigma$ is a time slice equals to the intersection of
asymptotical boundary and Cauchy surface in the bulk [5,6]. Action
for $WDW$ patch is given by the summation of the action and
 all boundary terms of this patch which is defined between the times $t_L$
and $t_R$ on the boundaries and at late time approximation could
be restricted by the Lloyd bound [7] as follows:
\begin{equation}
\frac{d(\mathcal{A}_{bulk}+\mathcal{A}_{boundary})}{dt}\leq2E,
\end{equation}
in which $E$ is the excited energy of the boundary quantum state.
It is now understood that the Lloyd`s bound does not need to hold
in holographic theories [8, 9]. To obtain the growth rate of
complexity on the boundary we must calculate the time derivative
of this action on the boundary by attention to conjecture of "CA"
given by (1.2) as
\begin{equation}
\frac{d\mathcal{C}}{dt}\leq\frac{2E}{\pi\hbar}.
\end{equation}
We can see the action growth rate by considering $CA$ conjecture
for the $WDW$ patch at late time approximation in $AdS$ black
holes is bounded as follows [5,6]:
\begin{gather}
\nonumber \frac{d\mathcal{A}}{dt} = 2M,\\
\nonumber \frac{d\mathcal{A}}{dt}\leq2[(M-\mu Q)-(M-\mu Q)_{gs}],\\
\frac{d\mathcal{A}}{dt}\leq2[(M-\Omega J)-(M-\Omega J)_{gs}],
\end{gather}
where $"gs"$ stands for ground state and the first equation
satisfies for neutral black hole and second and third ones happens
for charged and rotating black holes, separately.  For a charged
black hole solution this bound depends on the size of the black
hole
 [5,6] and can violate the Lloyd`s bound by assuming that it holds. The argument in the original paper
 [5]
involved the weak gravity conjecture. It is found in [10] that all
size of charged black hole violate the original bound (1.4). The
authors corrected the bound by investigating some other AdS black
holes like Kerr-AdS black hole and charged Gauss-Bonnet-AdS black
hole and summarized the final result as following bound:
\begin{equation}
\frac{d\mathcal{A}}{dt}\leq(M-\Omega J-\mu Q)_+-(M-\Omega J-\mu Q)_-,
\end{equation}
 at which $+$ and $-$ stand for the states of the most outer and inner horizon, respectively. Equality satisfies for stationary AdS black holes in Einstein gravity and charged
AdS black hole in Gauss-Bonnet gravity. In general non-stationary cases inequality would be expected. In the other words the work [10] shows that there is a universal formula for the
action growth of stationary black holes
 for which the Lloyd bound is independent of the charged black hole size and so would
  be satisfied for any arbitrary size of charged black hole.\\
   In this work we use non-rotating case of this universal form of the action
   growth and seek relation between this bound
 and the non-abelian charges. To do so we choose the Yang-Mills (YM) field given in the
    Maxwell-power-Yang-Mills theory propagating in AdS spacetime.  There are two important motivations which encourage
    us to consider YM field in our study: at first we can find YM equations in the low energy limit of some string theory models which leads
     to certain revisions of the no-hair theory of black hole physics. Secondly, the unification of general relativity and
     quantum mechanics in high energy regime of the most of string theory models is possible when it predicts a non-abelian gauge field.    \\
    In the other side, considering YM fields let us study small range effects inside the nuclei
    which are neglected in the long range effects of Maxwell field. Actually, these short-range effects are due to some
     length scales arisen by confinement in YM theory. Indeed, a confining YM theory undergoes a confinement-deconfinement phase transition
     in CFT side at a such given scale which leads to these short range effects. In the AdS/CFT dictionary, it is interesting we seek operators in the
field theory side because they correspond to other quantities
given in the gravity side. Actually in a top-down approach we can
specify dual field of the used gravity model which is included
both Maxwell and YM fields. It would be an interesting subject to
find familiar results from CFT side which come from the string
theory solutions. This can be considered as a future work which we
will study. Another important aspect of thermal systems is chaos
which could be described with its corresponding dual in the bulk
as the shock waves near the horizon of AdS black holes [11,12,13].
Actually a perturbation disturbs the geometry of the black hole
and then grows by time due to the backreaction effects. During the
chaos behaviors, the similar initial orthogonal quantum states are
changed to some totally different states. An interesting point
that makes the study of shock wave geometry important, is the
reflection of complexity on the boundary as the existence of a
firewall. When the perturbation on the boundary depends on
transverse coordinates then corresponding complexity is closely
connected to the speed of the perturbation spreading (butterfly
velocity) in spatial directions. This butterfly velocity is
studied by out-of-time order four-point function between pairs of
local operators $V(t=0)$ and $W(t)$ which are separated in spatial
coordinates such that [14] \begin{equation} \langle
V(0,x)W(t,y)V(0,x)W(t,y)\rangle_\beta,
\end{equation}
in which $\beta$ or the inverse of the temperature stands for
thermal expectation value. After the scrambling time $t_*$ the
butterfly effects could be seen by a sudden decay as follows [15].
\begin{equation}
\frac{\langle V(0,x)W(t,y)V(0,x)W(t,y)\rangle_\beta}{\langle V(0,x)V(0,x)\rangle_\beta\langle W(t,y)W(t,y)\rangle_\beta}\sim 1-
e^{\lambda_L\big(t-t_*+\frac{|x-y|}{v_B}\big)},
\end{equation}
where $\lambda_L=2\pi/\beta$ stands for the Lyapunov exponent and $v_B$ is the butterfly velocity. There is a wide
 variety of works has been devoted to the calculation of butterfly velocity for gravity theories in the bulk such as [16] for
  planar black hole in the Einstein's general theory of relativity framework, [15] for the topologically massive gravity (TMG) and the new massive
   gravity (NMG), [17] for the Einstein-Gauss-Bonnet gravity and other works [11,12,18,19].
The action growth could be affected in the shock wave geometry by
attention to the characteristics of gravity models. Authors of the
references  [20,21] used particular gravity models and showed that
the butterfly velocity is depended to change of the source fields
due to the alteration of the action growth. However there is no
fundamental connection in quantum information between the growth
of complexity and quantum chaos and the relations found in [20,21]
are just merely circumstantial and do not hold in general. In this
paper we devotes our work to the black hole solution including
both Maxwell
 and YM fields in the Einstein-Maxwell-power-Yang-Mills gravity. We seek the constraints and circumstances on the parameter of YM theory when the
  Lloyd bound is applied on the complexity growth rate.
The outline of this work is as follows: in section 2 we obtain the
evolution of complexity growth
 at the late time approximation and check the Lloyd bound in presence
 of the YM field and find a constraint condition on the parameters of the gravity theory.
 In section 3 we focus on the effects of a disturbance on the boundary
 and the spreading of shock wave in the used gravity model.
 We also discuss the effects of the parameters of this theory on the butterfly velocity.
  Last section denotes to summarize of the results and the conclusion.

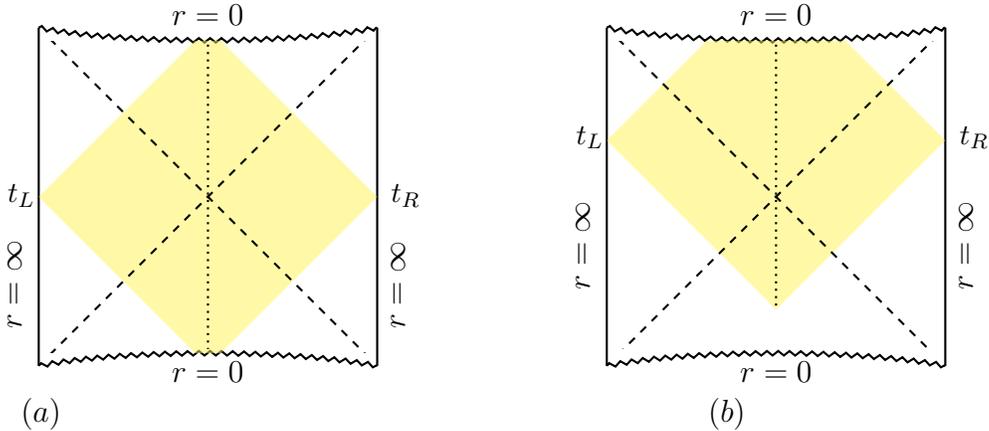
\begin{figure}[t]
\begin{center}
\begin{tabular}{cc}
\setlength{\unitlength}{1cm}
\begin{tikzpicture}[scale=1.5]
\draw
 [thick,decorate,decoration={zigzag,segment length=1.5mm,
amplitude=0.3mm}] (0,3) .. controls (.75,2.85) and (2.25,2.85) ..
(3,3);
 \draw [thick]  (0,0) -- (0,3); \draw [thick]  (3,0) --
(3,3);
 \draw [thick,decorate,decoration={zigzag,segment
length=1.5mm,amplitude=.3mm}]  (0,0) .. controls (.75,.15) and
(2.25,.15) .. (3,0); \clip (3,0.12) rectangle (0,2.88);
\fill[fill=yellow!80,opacity=0.5] (1.5,0) node {} -- (3,1.5) node
{} -- (1.5,3) node {} -- (0,1.5) node {}; \draw [thick,dashed]
(0,0) -- (3,3); \draw [thick,dashed]  (0,3) -- (3,0); \draw
[thick,dotted]  (1.5,2.85) -- (1.5,0.15);
\end{tikzpicture}
\qquad\qquad & \hspace{1.5cm}
\begin{tikzpicture}[scale=1.5]
\draw [thick,decorate,decoration={zigzag,segment length=1.5mm,
amplitude=0.3mm}] (0,3) .. controls (.75,2.85) and (2.25,2.85) ..
(3,3);
 \draw [thick]  (0,0) -- (0,3);
 \draw [thick]  (3,0) --
(3,3); \draw [thick,decorate,decoration={zigzag,segment
length=1.5mm,amplitude=.3mm}]  (0,0) .. controls (.75,.15) and
(2.25,.15) .. (3,0);
 \clip (3,0.15) rectangle (0,2.88);
\fill[fill=yellow!80,opacity=.5] (1.5,0.5) node {} -- (3,2.0) node
{} -- (1.5,3.5) node {} -- (0,2.0) node {}; \draw [thick,dashed]
(0,0) -- (3,3); \draw [thick,dashed]  (0,3) -- (3,0);
 \draw
[thick,dotted]  (1.5,2.85) -- (1.5,0.5);
\end{tikzpicture}
\qquad
  \put(-355,63){\small $t_{L}$ }
  \put(-210,63){\small $t_{R}$}
  \put(-355,15){\rotatebox{90}{$r = \infty$}}
  \put(-210,15){\rotatebox{90}{$r = \infty$}}
  \put(-140,85){\small $t_{L}$}
  \put(+5,85){\small $t_{R}$}
  \put(-140,30){\rotatebox{90}{$r = \infty$}}
  \put(+5,30){\rotatebox{ 90}{$r = \infty$}}
  \put(-293,130){$r =0$}
  \put(-293,-5){$r =0$}
  \put(-78,130){$r =0$}
  \put(-78,-5){$r =0$}
  \put(-350,-20){$(a)$}
  \put(-90,-20){$(b)$}
\end{tabular}
\end{center}
\caption{ Penrose diagram for a neutral two sided black hole and
WDW patch in ($a$) initial and ($b$) late time regimes. }
\label{fig-WDW}
\end{figure}

\begin{figure}[h]
\begin{center}
\begin{tabular}{cc}
\setlength{\unitlength}{1cm}

\begin{tikzpicture}[scale=1]

\draw [thick]  (0,0) -- (0,3); \draw
[thick,decorate,decoration={zigzag,segment
length=1.5mm,amplitude=.3mm}]  (0,3) -- (0,5); \draw [thick] (0,0)
-- (0,-1.5); \draw [thick]  (3,0) -- (3,3); \draw
[thick,decorate,decoration={zigzag,segment
length=1.5mm,amplitude=.3mm}]  (3,3) -- (3,5); \draw [thick] (3,0)
-- (3,-1.5);

\draw [thick,dotted]  (3,3) -- (.95,5); \draw [thick,dotted]
(0,3) -- (2.05,5); \draw [thick,dotted]  (0,0) -- (1.5,-1.5);
\draw [thick,dotted]  (3,0) -- (1.5,-1.5); \draw [thick,dotted]
(0,0) -- (-1.5,1.5); \draw [thick,dotted]  (-1.5,1.5) -- (0,3);
\draw [thick,dotted]  (3,0) -- (4.5,1.5); \draw [thick,dotted]
(4.5,1.5) -- (3,3); \draw [thick,dotted]  (0,3) -- (-1.5,4.5);
\draw [thick,dotted]  (-1.5,4.5) -- (-1,5); \draw [thick,dotted]
(3,3) -- (4.5,4.5); \draw [thick,dotted]  (4.5,4.5) -- (4,5);

\draw [thick,dotted]  (0,0) -- (-1.5,-1.5); \draw [thick,dotted]
(3,0) -- (4.5,-1.5);

\draw [thick,blue]  (1.5,3) -- (3,1.5); \draw [thick,blue]
(3,1.5) -- (1.5,0);

\clip (3,0) rectangle (0,6); \fill[fill=blue!50,opacity=.5]
(0.1,1.6) node {} -- (1.5,3) node {} -- (1.4,3.1) node {} --
(0,1.7) node {};

\draw [thick,blue]  (0,1.7) -- (1.4,3.1); \draw [thick,blue]
(1.4,3.1) -- (1.5,3);

\clip (3,0) rectangle (0,3); \fill[fill=blue!10,opacity=.5]
(1.6,.1) node {} -- (3,1.5) node {} -- (1.5,3) node {} --
(0.1,1.6) node {};

\clip (3,0) rectangle (0,3); \fill[fill=red!20,opacity=.5] (1.5,0)
node {} -- (1.6,.1) node {} -- (.1,1.6) node {} -- (0,1.5) node
{};

\draw [thick,red]  (1.5,0) -- (0,1.5); \draw [thick,red]  (1.5,0)
-- (1.6,0.1); \draw [thick,red]  (0,1.5) -- (0,1.5); \draw
[thick,red]  (0,1.5) -- (1.5,3);

\draw [thick,red]  (1.5,3) -- (3,1.5); \draw [thick,red]  (3,1.5)
-- (1.5,0);

\draw [thick,blue]  (1.6,0.1) -- (0,1.7);

\draw [thick,violet]  (1.6,.1) -- (3,1.5); \draw [thick,violet]
(3,1.5) -- (1.5,3);

\draw [thick,dotted]  (3,0) -- (0,3); \draw [thick,dotted]  (0,0)
-- (3,3);

\end{tikzpicture}
\qquad\qquad & \hspace{1.5cm} \qquad
  \put(-188,82){\scriptsize $t_{L}\rightarrow$ }
  \put(-202,90){\scriptsize $t_{L}+\epsilon\rightarrow$ }
  \put(-82,84){\scriptsize $\leftarrow t_{R}$ }
  \put(-143,74){\rotatebox{45}{$r_2$}}
  \put(-118,80){\rotatebox{-45}{$r_2$}}
  \put(-143,107){\rotatebox{-45}{$r_2$}}
  \put(-122,95){\rotatebox{45}{$r_2$}}
  \put(-148,142){\rotatebox{45}{$r_{1}$}}
  \put(-112,146){\rotatebox{-45}{$r_1$}}
  \put(-146,22){\rotatebox{-45}{$r_0$}}
  \put(-114,19){\rotatebox{45}{$r_0$}}
  \put(-200,150){\rotatebox{-45}{$r_\infty$}}
  \put(-54,149){\rotatebox{45}{$r_\infty$}}
  \put(-190,15){\rotatebox{45}{$r_4$}}
  \put(-73,20){\rotatebox{-45}{$r_4$}}
  \put(-200,107){\rotatebox{45}{$r_3$}}
  \put(-200,62){\rotatebox{-45}{$r_3$}}
  \put(-65,115){\rotatebox{-45}{$r_3$}}
  \put(-65,52){\rotatebox{45}{$r_3$}}

\end{tabular}
\end{center}
\caption{ Penrose diagram for a charged two sided black hole with
multiple horizons and WDW patch in
 late time approximation. $r_1$ is the most internal horizon and $r_0$ is the null spatial infinity. The wavy
  lines indicate the singularities at $r=0$, and $r_\infty$ stands for $r=-\infty$. }
\label{fig-WDW}
\end{figure}
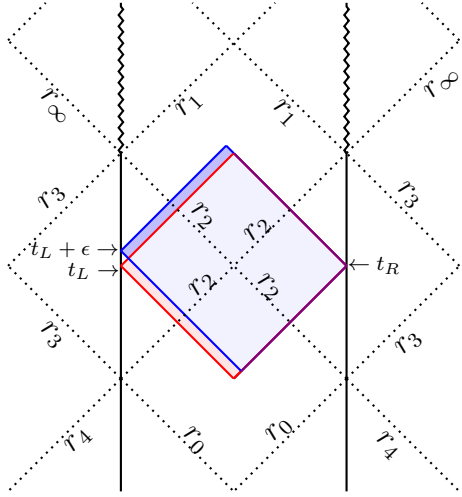

\section{The complexity growth}
As we know from [5,6] the growth rate of the action of a WDW
patch of the two-sided black hole at the late time, i.e.
 $t_L+t_R>> \beta$, corresponds to the increasing rate of complexity of the boundary state. At the late time and without any shock wave the
  contribution of the region behind the past horizon goes to zero exponentially. By adding any kind of conserved charges WDW patch terminates
   slower than neutral case and Lloyd bound must be generalized due to changing of average energy of the quantum state related to the ground state.
    In figure 1 we can see the general Penrose diagram and WDW patch for a neutral two sided black hole for initial
    times and late times approximations. In figure ($1.a$) which is indicated initial times the patch intersects both the future and past singularities
    at $r=0$, but in late time indicated by figure ($2.a$) only intersects the future one.
In this section we are about to consider a Maxwell-Yang-Mills
theory for the black hole inside the bulk and study action growth
rate for the late time approximation. Also we study the Lloyd
bound in presence of the conserved charges of this model. The
action for the Einstein-Maxwell-power-Yang-Mills gravity with a
negative cosmological constant in 4-dimension is given by [22,23]:
\begin{gather}
\nonumber \mathcal{A}=\frac{1}{16\pi G}\int d^4x\sqrt{-g}\big(\mathcal{R}+\frac{6}{\ell^2}-F_{\mu\nu}F^{\mu\nu}-[\mathbf{Tr}(F_
{\mu\nu}^{(a)}F^{(a)\mu\nu})]^\gamma\big)\\
+\frac{1}{8\pi G}\int_{\partial\mathcal{M}} d^3x\sqrt{-h}\mathcal{K},
\end{gather}
in which the first integral equation represents the action in the
bulk and the
 second integral equation is the boundary part of WDW patch (WDW patch located
     in our two sided black hole is indicated in figure 2 in which only dark blue region contributes to the complexity growth at late time
     approximation.). Radius of AdS spacetime is indicated by $\ell$ and $\mathcal{R}$ stands for Ricci scalar.
 $\gamma$ is a real positive parameter and the  Yang-Mills tensor fields  $F_{\mu\nu}^{(a)}$ are defined as follows.
\begin{equation}
F_{\mu\nu}^{(a)}=\partial_\mu A_\nu^{(a)}-\partial_\nu A_\mu^{(a)}+\frac{1}{2\sigma}C^{(a)}_{(b)(c)}A^b_\mu A^c_\nu,
\end{equation}
in which $\sigma$ is a coupling constant and $C^{(a)}_{(b)(c)}$
are the structure constants of $(d-1)(d-2)/2$ parameter Lie group
$G$ in general $d$-dimensional theory and $A_\mu^{(a)}$ is the
$SO(d-1)$ gauge group Yang-Mills potentials. According to Wu-Yang
ansatz the YM invariant $\mathcal{F}$ reduces to the following
form [23].
\begin{equation}
\mathcal{F}_{YM}=\mathbf{Tr}(F_{\mu\nu}^{(a)}F^{(a)\mu\nu})=-\frac{2q_{YM}^2}{r^4}.
\end{equation}
The electromagnetic tensor field is defined by the usual Maxwell
potential $A_\mu$ such that
\begin{equation}
F_{\mu\nu}=\partial_\mu A_\nu-\partial_\nu A_\mu,
\end{equation}
for which the gauge invariant counterpart $\mathcal{F}_{EM}$ is
\begin{equation}
\mathcal{F}_{EM}=F_{\mu\nu}F^{\mu\nu}=-\frac{2q_{E}^2}{r^4}.
\end{equation}
For a spherically symmetric 4 dimensional static space time metric
equation defined in a Schwarzschild frame is given by
\begin{equation}
ds^2=-f(r)dt^2+\frac{dr^2}{f(r)}+r^2{d\Omega_2}^2,
\end{equation}
which by substituting it into the Einstein metric equation
obtained from (2.1) one can infer that there is a black hole
solution [24] as \begin{equation}
f(r)=1-\frac{2M}{r}+\frac{r^2}{\ell^2}+\frac{q_E^2}{r^2}+\frac{Q}{r^{4\gamma-2}},~~~Q=\frac{2^{\gamma-1}}{4\gamma-3}q_{YM}^{2\gamma}.
\end{equation} for $\gamma\neq\frac{3}{4}$ and \begin{equation}
f(r)=1-\frac{2M}{r}+\frac{r^2}{\ell^2}+\frac{q_E^2}{r^2}-\frac{Q_0lnr}{r},~~~Q_0=2^{-\frac{1}{4}}q_{YM}^{\frac{3}{2}}.
\end{equation}
for  $\gamma=\frac{3}{4}$ where $q_{EM}$ and $q_{YM}$ correspond to the electric and Yang-Mills charges respectively.
 Writing $f(r)$ like an equipotential surface $f(r)=constant$ the
 first law of the black hole thermodynamics could be derived  for which
\begin{gather}
\nonumber df(S,M,P,q_E,q_{YM})=0\\=\frac{\partial f}{\partial
S}dS+\frac{\partial f}{\partial M}dM+\frac{\partial f}{\partial
P}dP +\frac{\partial f}{\partial q_E}dq_E+\frac{\partial
f}{\partial q_{YM}}dq_{YM},
\end{gather}
where $S=\pi r^2$ is the entropy of the black hole and
$P=\frac{3}{8\pi\ell^2}$ is the pressure of AdS spacetime.
Applying the above relation we can obtain
\begin{equation}
dM=TdS+VdP+\phi_Edq_E+\phi_{YM}dq_{YM},
\end{equation}
where $T=\frac{1}{2r}\partial_Sf$ is the temperature, $V=4\pi
r^3/3$ is the thermodynamic volume, $\phi_E=q_E/r$ stands for
electric potential and the Yang-Mills potential reads.
\begin{gather}
\nonumber\phi_{YM}=\bigg(\frac{\gamma}{4\gamma-3}\bigg)\frac{2^{\gamma-1}q_{YM}^{2\gamma-1}}{r^{4\gamma-3}},~~
\textit{for}~~\gamma\neq\frac{3}{4}\\
\phi_{YM}=-3\times2^{-\frac{9}{4}}q_{YM}^{\frac{1}{2}}lnr,~~\textit{for}~~\gamma=\frac{3}{4}.
\end{gather}
Regarding the above solution, Ricci scalar could be achieved as:
\begin{equation}
\mathcal{R}=-\frac{12}{\ell^2}-\frac{4Q}{r^{4\gamma}}(4\gamma^2-7\gamma+3)~~\textit{for}~~\gamma\neq\frac{3}{4},
\end{equation}
and
\begin{equation}
\mathcal{R}=-\frac{12}{\ell^2}+\frac{Q_0}{r^3}~~\textit{for}~~\gamma=\frac{3}{4}.
\end{equation}
Therefore, the growth rate of the bulk action given by first
integral equation in (2.1), can be calculated at late time
approximation as follows:
\begin{gather}
\nonumber\frac{d\mathcal{A}_{bk}}{dt}=\frac{1}{16\pi
G}\int\int_{r_-}^{r_+}r^2\bigg[-\frac{6}{\ell^2}-\frac{4Q}{r^{4\gamma}}(4\gamma^2-7\gamma+3)+\frac{2q_E^2}{r^4}
+\frac{2^\gamma q_{YM}^{2\gamma}}{r^{4\gamma}}\bigg]drd\Omega_2\\
\nonumber=-\frac{1}{2\ell^2}(r_+^3-r_-^3)-\frac{q_E^2}{2}(\frac{1}{r_+}-\frac{1}{r_-})+\bigg(\frac{2\gamma-3}{4\gamma-3}\bigg)2^{\gamma-2}
q_{YM}^{2\gamma}
(\frac{1}{r_+^{4\gamma-3}}-\frac{1}{r_-^{4\gamma-3}}),\\
\end{gather}
for $\gamma\neq\frac{3}{4}$ and
\begin{gather}
\nonumber\frac{d\mathcal{A}_{bk}}{dt}=\frac{1}{16\pi
G}\int\int_{r_-}^{r_+}r^2\bigg[-\frac{6}{\ell^2}+\frac{Q_0}{r^3}+\frac{2q_E^2}{r^4}
+\frac{2^{3/4} q_{YM}^{3/2}}{r^{3}}\bigg]drd\Omega_2\\
\nonumber=-\frac{1}{2\ell^2}(r_+^3-r_-^3)-\frac{q_E^2}{2}(\frac{1}{r_+}-\frac{1}{r_-})+3\times2^{-9/4}q_{YM}^{3/2}ln\big(\frac{r_+}{r_-}\big),\\
\end{gather}
for $\gamma=\frac{3}{4}$ respectively. In the above integral
equations we put $\Omega_2/4\pi G=1$. It must be noted that
 spatial integral is calculated between two horizons $r_+$ and $r_-$ which are the outer and inner horizons of the black hole. As we can see
 from (2.7) there are multiple horizons which are obtained from $f(r)=0$ and its number depends on $\gamma$. So for any values of $\gamma$
 we will have several horizons (see eq. (2.8) as a special case), but $r_-$ and $r_+$ would be the most internal and outer horizons,
  respectively. Penrose diagram for a black hole solution depends on the number of horizons as it is shown for a
  black hole with multiple horizons in ref.
  [25]. In our case for any values of $\gamma$ we can have different real roots obtained from $f(r)=0$, but in a general form the Penrose diagram looks
   like figure 2 in which $r_1$ is the most internal horizon and so $0<r_-\equiv r_1<r_2<r_3<...<r_+$, also $r_\infty$ stands for $r=-\infty$
    and $r_0$ indicates spatial null infinity. One can see the position of singularities and multiple horizons in ref. [25]. As we mentioned earlier
     figure 2 deoicted WDW patch at late time approximation in our multiple horizon case.
In the other side, the boundary part (second integral equation in
eq. (2.1)) of the action growth rate at late time approximation is
given by:
\begin{gather}
  \frac{d\mathcal{A}_{bd}}{dt}=\frac{1}{8\pi G}\int_{\partial\mathcal{M}}{d\Omega_2}(\sqrt{-h}\mathcal{K}) =\frac{1}{2}\bigg[r^2\sqrt{f(r)}
  \bigg(\frac{2}{r}\sqrt{f(r)}+\frac{f^{\prime}(r)}{2\sqrt{f(r)}}\bigg)\bigg]_{\partial\mathcal{M}},
\end{gather}
where the extrinsic curvature is defined by
\begin{equation}
\mathcal{K}=\frac{1}{r^2}\frac{\partial}{\partial r}(r^2\sqrt{f(r)})=\frac{2}{r}\sqrt{f(r)}+\frac{f^{\prime}(r)}{2\sqrt{f(r)}}.
\end{equation}
 One can see that the Eq. (2.16)
only contains the Gibbons-Hawking term. However, for the
computation of complexity one needs further boundary terms for the
abelian and non-abelian fields as well as extra terms for the null
boundaries and corners of the WDW patch. In fact these
extra terms become negligible just at late time approximation
which we considered here. So by attention to the metric solutions
(2.7) and (2.8) we can obtain:
\begin{gather}
 \nonumber \frac{d\mathcal{A}_{bd}}{dt}=(r_+-r_-)+\frac{3}{2\ell^2}(r_+^3-r_-^3)+\frac{q_E^2}{2}\big(\frac{1}{r_+}
  -\frac{1}{r_-}\big)\\
  +\frac{(3-2\gamma)Q}{2}\big(\frac{1}{r_+^{4\gamma-3}}-\frac{1}{r_-^{4\gamma-3}}\big),
\end{gather}
for $\gamma\neq\frac{3}{4}$ and
\begin{gather}
 \nonumber \frac{d\mathcal{A}_{bd}}{dt}=(r_+-r_-)+\frac{3}{2\ell^2}(r_+^3-r_-^3)+\frac{q_E^2}{2}\big(\frac{1}{r_+}
  -\frac{1}{r_-}\big)
  -3\times2^{-9/4}q_{YM}^{3/2}ln\big(\frac{r_+}{r_-}\big),
\end{gather}
for $\gamma\neq\frac{3}{4}$. Hence the total growth rate of the
action for all values of $\gamma$ is achieved as follows.
\begin{equation}
\frac{d\mathcal{A}}{dt}=(r_+-r_-)+\frac{r_+^3-r_-^3}{\ell^2}.
\end{equation}
It is useful to rewrite the total growth action equation with
respect to the black hole characteristics like charges and mass.
By using the horizon equations $f(r_+)=f(r_-)=0,$ one can obtain
the following relations for the electric charge and mass of the
black hole.
\begin{equation}
q_E^2=r_+r_-\bigg[1+\frac{1}{\ell^2}\bigg(\frac{r_+^3-r_-^3}{r_+-r_-}\bigg)+\frac{2^{\gamma-1}}{4\gamma-3}q_{YM}^{2\gamma}
\bigg(\frac{r_+^{-(4\gamma-3)}-r_-^{-(4\gamma-3)}}{r_+-r_-}\bigg)\bigg],
\end{equation}
with
\begin{equation}
M=\frac{1}{2}\Bigg[(r_++r_-)+\frac{1}{\ell^2}\bigg(\frac{r_+^4-r_-^4}{r_+-r_-}\bigg)
+\frac{2^{\gamma-1}}{4\gamma-3}q_{YM}^{2\gamma}\bigg(\frac{r_+^{-4(\gamma-1)}-r_-^{-4(\gamma-1)}}{r_+-r_-}\bigg)\Bigg],
\end{equation}
for $\gamma\neq\frac{3}{4}$ and,
\begin{equation}
q_E^2=r_+r_-\bigg[1+\frac{1}{\ell^2}\bigg(\frac{r_+^3-r_-^3}{r_+-r_-}\bigg)-2^{-1/4}q_{YM}^{3/2}\bigg(\frac{ln(\frac{r_+}{r_-})}{r_+-r_-}\bigg)\bigg],
\end{equation}
with
\begin{equation}
M=\frac{1}{2}\Bigg[(r_++r_-)+\frac{1}{\ell^2}\bigg(\frac{r_+^4-r_-^4}{r_+-r_-}\bigg)
+2^{-1/4}q_{YM}^{3/2}\bigg(\frac{r_+lnr_+-2r_+lnr_-+r_-lnr_-}{r_+-r_-}\bigg)\Bigg].
\end{equation}
for $\gamma=\frac{3}{4}$ respectively. By attention to these
definitions one could rewrite the total action growth rate for
various values of $\gamma$ as follows:
\begin{equation}
  \frac{d\mathcal{A}}{dt}=-q_E^2\big(\frac{1}{r_+}-\frac{1}{r_-}\big)
  -\bigg(\frac{2^{\gamma-1}}{4\gamma-3}\bigg)q_{YM}^{2\gamma}\bigg(\frac{1}{r_+^{4\gamma-3}}-\frac{1}{r_-^{4\gamma-3}}\bigg).
\end{equation}
for $\gamma\neq\frac{3}{4}$ and
\begin{equation}
  \frac{d\mathcal{A}}{dt}=-q_E^2\big(\frac{1}{r_+}-\frac{1}{r_-}\big)
  +2^{-\frac{1}{4}}q_{YM}^{\frac{3}{2}}ln\big(\frac{r_+}{r_-}\big).
\end{equation}
for $\gamma=\frac{3}{4}$ respectively. So by attention to the
conjugated potentials $(\phi_{E}, \phi_{YM})$ which are derived
earlier in this section, the total action growth at the late time
approximation would be:
\begin{equation}
\frac{d\mathcal{A}}{dt}=(M-\phi_{E+}q_E-\frac{1}{\gamma}\phi_{YM+}q_{YM})-(M-\phi_{E-}q_E-\frac{1}{\gamma}\phi_{YM-}q_{YM}),
\end{equation}
where $\gamma$ can be take all real values. It is simple to check
the Lloyd bound is satisfied for $\gamma\geq1$ [7] regarding to the
equation (1.4) and so the case where $\gamma=\frac{3}{4}$, however all
situations where $\gamma<1$ violate the Lloyd bound.

\section{The complexity growth in a shock wave geometry}
In this section we are about to study the above mentioned problem
but in presence of a chaotic shock wave which makes perturbed the
background geometry. Actually when a shock wave is sent into the
bulk at time $t_w$, a precursor operator $W(t)$ acts on the
boundary at the same time. It will be effective on the initial
state of black hole which is a thermofield double state ($TFD$),
so the old state changes to $W(t_w)|TFD\rangle$. For a local shock
wave this operator depends on transverse coordinates and localizes
on the boundary at $x$. $W(t_w,x)$ grows in this spatial direction
vs the time which leads to the growth of action due to the
perturbation on the boundary. In the other word action growth
depends on the growth velocity of perturbation on the boundary in
spatial direction which is called "butterfly velocity". To see the
evolution of action growth in the presence of a local shock wave
it would be useful to study this perturbation in more details.
Shock wave perturbs the black hole solution by injection of a
small amount of energy from the boundary of AdS spacetime towards
the horizon. This perturbation grows by raising the time due to
the back reaction effects and so propagates on the horizon. By
attention to the work  presented by Dary and t`Hooft
 [26], we study the problem in Kruskal null coordinates ($u,v$) as ,
\begin{equation}
uv=-\text{exp}[\frac{4\pi}{\beta}r_*],~~~u/v=-\text{exp}[-\frac{4\pi}{\beta}t],
\end{equation}
in which $\beta$ is proportional to the inverse of Hawking
temperature and $r_*$ is a function of $r$ which is defined by
$dr_*=\frac{dr}{f(r)}$. The effect of shock wave geometry is
considered as the effect of a massless particle at $u=0$ which
moves in the direction of $v$ with the speed of light. So geometry
for $u<0$ stays unchanged like (2.6) and in Kruskal form will be
[20] :
\begin{equation}
ds^2=-2A(u,v)dudv+B(u,v){d\vec{x}_2}^2,
\end{equation}
where
\begin{equation}
A(u,v)=-\frac{4}{uv}\frac{f(r)}{[f^{\prime}(r_h)]^2},~~~B(u,v)=r^2
\end{equation}
while for $u>0$ the particle moves in direction of shifted advance
coordinate $v\rightarrow v+\alpha(x)$, where $\alpha(x)$ is the
shift function. So in general for all values of $u$ we can
determine new coordinates system from the old ones by using the
well known step function $\theta(u)$ such as follows.
\begin{gather}
  \nonumber \hat{u}\equiv u \\
  \nonumber \hat{v}\equiv v+\theta(u)\alpha(x)\\
  \hat{x}\equiv x.
\end{gather}
By these transformations the metric and the energy momentum tensor
are affected. If there are some stress tensor of matter fields
then their non-zero components have changed to new form. This new
geometry and new energy momentum tensor still satisfies the
Einstein equation, $\mathcal{\hat{G}}=\mathcal{\hat{T}}_{matter}$,
in which $\mathcal{\hat{G}}$ and $\mathcal{\hat{T}}_{matter}$ are
the Einstein tensor and the energy
 momentum tensor of all matter fields defined in the new coordinates
 system, respectively.
  After acting a scalar operator at $t_w<0$ and
 producing the shock wave, this perturbation propagates along $\hat{u}=0$ and its stress-energy tensor will have only $\hat{u}\hat{u}$ component [6],
\begin{equation}
\mathcal{\hat{T}}_{\{shock\}\hat{u}\hat{u}}\sim\delta(\hat{u})\exp[\frac{2\pi|t_w|}{\beta}].
\end{equation}
By adding this part of perturbation to the stress-energy tensor
and solving the Einstein equation we find a relationship for shift
function $\alpha(x)$. If we consider this function independent of
the transverse coordinates, $\theta$ and $\phi$ which are valid
just for spherical shock waves, so it will be obtained as:
\begin{equation}
\alpha\sim e^{\lambda_L(|t_w|-t_*)},
\end{equation}
in which $\lambda_L=\frac{2\pi}{\beta}$ is the Lyapunov exponent
and the scrambling time $t_*=\frac{\beta}{2\pi}ln(S)$ is related
to the entropy $S$ and it is a delay time on the action growth due
to the "switchback effect". In other side when the shock wave is
local, the shift function depends on the transverse coordinates.
By solving the equations in this case we have an extra term in the
above exponential part. If there is just one transverse
coordinate, named $x$, so the shift function yields:
\begin{equation}
\alpha(x)\sim e^{\lambda_L(|t_w|-t_*-\frac{|x|}{v_B})},
\end{equation}
where \begin{equation} v_B=\sqrt{\frac{f^{\prime}(r_h)}{4r_h}},
\end{equation}
is called the butterfly velocity and presents the speed of the
local shock wave on the boundary. $r_h$
is the outer horizon radius which is achieved by $f(r_h)=0$.\\
The action behind the future horizon is
$A_{future}=2M\lambda_{L}\int \ln(u_0v_R)dx$ where $v_R$  is the
right boundary of the WDW patch, $v_R=v_0+\alpha(x)$ (see [21])
which by substituting the shift function (3.7) reads
\begin{equation}
\mathcal{A}_{future}=\frac{2M}{L\lambda_L}\int\ln e^{\lambda_L(|t_w|-t_*+t_L-\frac{|x|}{v_B})}dx
\end{equation}
where $L=\int dx$ is the length of the transverse direction that
goes to infinite for a planar black hole [20].  Similarly,
substituting the expression of the shift equation (3.7) into the
action behind the past horizon, we obtain
 \begin{equation}
\mathcal{A}_{past}=\frac{2M}{L\lambda_L}\int\ln
e^{\lambda_L(|t_w|-t_*-t_R-\frac{|x|}{v_B})}dx
\end{equation} where
the upper limit of the integrals should be chosen as
$|x|=v_B(|t_w|-t_*-t_R)$ called as maximal transverse coordinate
coming from "the large shift condition" at which shock wave could
be effective and described through $|t_w|-t_*-\frac{|x|}{v_B}\geq
t_{R}$. Now we can obtain action of WDW patch by adding (3.9) and
(3.10) as follows.
\begin{equation}
\mathcal{A}_{WDW}=\mathcal{A}_{future}+\mathcal{A}_{past}=2M(t_L+t_R)+2Mv_B(|t_w|-t_*-t_R)^2.
\end{equation}
The first term denotes action of WDW patch for the case with no
shock wave and the second term that depends on the butterfly
velocity represents the effect of local shock wave. As we can see
the second term only depends on $t_R$ because the shock wave
reaches the right side of our two sided black hole. By keeping one
of boundary times fixed and varying it with respect to another
time we can study the growth rate of boundary complexity. As we
can see action growth with respect to $t_L$ leads to the same
result with no shock wave case, but it linearly depends on
butterfly velocity when it is calculated with respect to $t_R$ or
$t_w$.
 Therefore to
study the action growth in the presence of a local shock wave it
will be necessary computing butterfly velocity and knowing how it
changes in various models of gravity. In the model under
consideration with $f(r)$ defined in (2.7) or (2.8) the butterfly
velocity reads
\begin{equation}
v_B=\frac{1}{2}\sqrt{\frac{1}{r_h^2}+\frac{3}{\ell^2}-\frac{q_E^2}{r_h^4}-\frac{1}{2}\bigg(\frac{2q_{YM}^2}{r_h^4}\bigg)^\gamma}.
\end{equation}
 As one can see the butterfly velocity in the presence of Yang-Mills fields takes an extra term which depends on the Yang-Mills charge $q_{YM}$
 and $\gamma$. It should be useful to compare $v_B (q_{YM}\neq0)$ with $v_B(q_{YM}=0)$.
  To do so we ignore
 the effect of Maxwell fields by setting $q_E=0$ for simplicity.  In the spacetime without the
electric and Yang-Mills charge we have simply the Schwarzschild
spacetime with its own single event horizon
$\textit{\textbf{r}}_h=2M$ and the
 butterfly velocity become simplified as
\begin{equation}
\textit{\textbf{v}}_B=\frac{1}{2}\sqrt{\frac{1}{{\textit{\textbf{r}}_h}^2}+\frac{3}{\ell^2}}.
\end{equation}
By attention to (2.7) and (2.8) for $q_E=0$ with fixed mass we
have $f(r)>{\textit{\textbf{f}}}~(r)$, in which
$\textit{\textbf{f}}~(r)\approx1-\frac{2M}{r}$ corresponds to the
Schwarzschild metric potential. This inequality is true for all
radiuses such as the event horizon
 $r_h$, so:
\begin{equation}
f(r_h)=0>{\textit{\textbf{f}}}~(\textit{r}_h),
\end{equation}
hence $\textit{\textbf{f}}~(\textit{r}_h)<0$ and since $\textit{\textbf{f}}~(\textit{\textbf{r}}_h)=0$
 as well, then we lead to the following statement.
\begin{equation}
\textit{\textbf{f}}~(\textit{r}_h)<\textit{\textbf{f}}~(\textit{\textbf{r}}_h)~\Rightarrow~\textit{r}_h<\textit{\textbf{r}}_h.
\end{equation}
It is easy to check from the above statement that for the
butterfly velocity in presence and absence of the YM fields one
can infer
\begin{equation}
v_B>\textit{\textbf{v}}_B.
\end{equation}
 Also it is interesting
 to know that the value of butterfly velocity is decreased by increasing $\gamma.$ Regarding to the results of these two
  sections we can conclude that the butterfly velocity has a same behaviour for all $\gamma$ and decreases by the increasing of $\gamma$, but the
  system violates the Lloyd's bound for $\gamma<1$ at late time approximations. We can see same situation in the gravity dual of a non-local theory
  in [28] in which the violation is correlated to some violation of the butterfly velocity studied in [29].

\section{Conclusion and summary}
In this work we used a black hole metric solution containing the
electric and the Yang-Mills charges and calculate corresponding
complexity growth rate by applying conjecture of
"complexity=action" [5,6]. We obtained that the Lloyd bound is
saturated only for $\gamma\geq1$ in late time approximation, but
not for values less than one. In the other side, when the boundary
is disturbed by a small amount of energy and so the spacetime
takes form of a shock wave geometry [4,10], then the spreading of
perturbation near the horizon affects on the complexity growth
rate  via the butterfly velocity. We show that the existence of
the Yang-Mills field causes to increase the butterfly velocity and
it decreases by raising the  $\gamma$ factor of the YM field. This
is in an opposition direction of [29] at which the violation of
Lloyd bound is correlated to the exceeding of butterfly velocity
from the speed of light.
 It is shown that in large shift
condition the action of WDW patch raises as linearly by increasing
the butterfly velocity $v_B$.

\section {Acknowledgment}
 Authors should thank to the editor and anonymous referees for their comments and suggestions which cause to improve this
 work for readers. They  (E. Y. and M. F.) have appreciate also for hospitality and generosity behavior of the Lorentz Institute of
 theoretical physics at Leiden University of Netherlands.

  \vskip .5cm
 \noindent
  {\bf References}
\begin{description}
\item[1.] L. Susskind, " Computational Complexity and Black Hole Horizons, hep-th/1402.5674.
\item[2.] L. Susskind, "Addendum to Computational Complexity and Black Hole Horizons, hep-th/1403.5695.
\item[3.] P. Hayden and J. Preskill, "Black holes as mirrors: Quantum information in random subsystems", JHEP 0709, 120 (2007)
hep-th/0708.4025.
\item[4.] D. Stanford and L. Susskind, "Complexity and Shock Wave Geometries", Phys. Rev. D 90, 126007 (2014)
hep-th/1406.2678.
\item[5.] A. R. Brown, D. A. Roberts, L. Susskind, B. Swingle and Y. Zhao, " Holographic Complexity Equals Bulk Action?", Phys. Rev. Lett. 116, 191301, (2016).
\item[6.] A. R. Brown, D. A. Roberts, L. Susskind, B. Swingle and Y. Zhao, "Complexity, action, and black holes " Phys. Rev. D 93, 086006, (2016).
\item[7.] S. Lloyd, "Ultimate physical limits to computation" Nature 406, 1047 (2000).
\item[8.]D. Carmi, S. Chapman, H. Marrochio, R. C. Myers and S.
Sugishita, "On the Time Dependence of Holographic Complexity",
JHEP, 11, 188, (2.17); hep-th/1709.10184
\item[9.] W. Cottrell and
M. Montero, "Complexity is Simple", JHEP,02, 039 (2017);
hep-th/1710.01175

\item[10.] Cai, Rong-Gen, Shan-Ming Ruan, Shao-Jiang Wang, Run-Qiu Yang, and Rong-Hui Peng. "Action growth for AdS black holes."
JHEP 2016, 161, (2016).
\item[11.] S.H. Shenker and D. Stanford, "Black holes and the butterfly effect", JHEP 1403, 067 (2014);
hep-th/1306.0622.
\item[12.] S.H. Shenker and D. Stanford, "Multiple shocks", JHEP 1412, 046 (2014); hep-th/1312.3296.
\item[13.] D.A. Roberts, D. Stanford and L. Susskind, "Localized shocks", JHEP 1503, 051 (2015); hep-th/1409.8180.
\item[14.] E. Perlmutter, "Bounding the space of holographic CFTs with chaos", JHEP 10, 069 (2016).
\item[15.] M. Alishahiha, A. Davody, A. Naseh, and S. F. Taghavi, "On butterfly effect in higher derivative gravities",
JHEP 11, 032, (2016);   hep-th/1610.02890.
\item[16.] X. H. Feng and H. Lu, "Butterfly Velocity Bound and Reverse Isoperimetric Inequality," Phys. Rev. D 95, 066001,
 (2017); hep-th/1701.05204.
\item[17.] W. H. Huang, "Holographic Butterfly Velocities in Brane Geometry and Einstein-Gauss-Bonnet Gravity with
Matters,", Phys. Rev. D 97, 066020 (2018); hep-th/1710.05765.
\item[18.] Y. Ling, P. Liu and J. P. Wu, "Note on the butterfly effect in holographic superconductor models", Phys. Lett. B 768, 288
(2017) hep-th/1610.07146.
\item[19.] R. G. Cai, X. X. Zeng and H. Q. Zhang, "Influence of inhomogeneities on holographic mutual information and butterfly
effect,",    JHEP 1707, 082, (2017); hep-th/1704.03989.
\item[20.] Y. G. Miao and L. Zhao, "Complexity/Action duality of the shock wave geometry in a massive gravity theory:,
 Phys. Rev. D 97, 024035 (2018);
hep-th/1708.01779.
\item[21.] S. A. Hosseini Mansoori and M. M. Qaemmaqami, "Complexity Growth, Butterfly Velocity and Black hole
Thermodynamics" hep-th/1711.09749.
\item[22.] H. El Moumni. "Revisiting the phase transition of AdS-Maxwell-power-Yang-Mills black holes via AdS/CFT tools", Phys. Lett., B776, 124, (2018).
\item[23.] S. H. Mazharimousavi, M. Halilsoy and Z. Amirabi, "Higher-dimensional thin-shell wormholes in
 Einstein–Yang–Mills–Gauss–Bonnet gravity" Class. Quantum Gravit 28, 025004 (2011).
\item[24.] M. Zhang, Z. Ying Yang, D. C. Zou, W. Xu and R. H. Yue , "P-V criticality of AdS black hole in the Einstein-Maxwell-power-Yang-Mills
gravity", Gen. Rel. Grav. 47, 14, (2015).
\item[25.] Gao, Changjun, Youjun Lu, Shuang Yu, and You-Gen Shen. "Black hole and cosmos with multiple horizons and multiple singularities
 in vector-tensor theories." Phys. Rev. D 97, 104013, (2018).
\item[26.] T. Dray and G.'t Hooft, "The gravitational shock wave of a massless particle", Nucl. Phys. B253, 173 (1985).
\item[27.] L. Lehner, R. C. Myers, E. Poisson, and R. D. Sorkin,
"Gravitational action with null boundaries", Phys. Rev. D 94,
084046 (2016); hep-th/1609.00207
\item[28.] C. Josiah, S. Eccles, W. Fischler and M. L. Xiao.
"Holographic complexity and noncommutative gauge theory" JHEP
2018, 3, 108,  (2018); hep-th/1710.07833.
\item[29.] W. Fischler, V. Jahnke and J. F. Pedraza,
 "Chaos and entanglement spreading in a non-commutative gauge theory" hep-th/1808.10050 (2018).

\end{description}

\end{document}